\documentclass{article}
\font\cero=cmss10 scaled 1728 
\usepackage{tensor}
\usepackage{verbatim}
\usepackage{amsfonts}
\usepackage{graphicx}
\usepackage{amssymb}
\usepackage{float}
\begin{document}
\begin{flushleft}
{\cero Hyperbolic ring based formulation for thermo field dynamics, quantum dissipation, entanglement, and holography}\\
\end{flushleft} 
{\sf R. Cartas-Fuentevilla,  J. Berra-Montiel, and O. Meza-Aldama}\\
{\it Instituto de F\'{\i}sica, Universidad Aut\'onoma de Puebla,
Apartado postal J-48 72570 Puebla Pue., M\'exico.} \\
{\it Facultad de Ciencias, Universidad Aut\'onoma de San Luis Potos\'i
Campus Pedregal, Av. Parque Chapultepec 1610, Col. Privadas del Pedregal, San
Luis Potos\'i, SLP, 78217, M\'exico.}\\ 
{\it Lux Systems, C.P. 27010, Blvrd Independencia 2002, Colonia Estrella, Torre\'on, Coah., M\'exico.}\\

ABSTRACT: 
The classical and quantum formulations for open systems related to dissipative dynamics are constructed on a complex hyperbolic ring, following universal symmetry principles, and considering the double thermal fields approach for modeling the system of interest, and the environment. The hyperbolic rotations are revealed as an underlying internal symmetry for the dissipative dynamics, and a chemical potential is identified as conjugate variable to the charge operator, and thus a grand partition function is constructed.
As opposed to the standard scheme, there are not patologies associated with 
the existence of many unitarity inequivalent representations on the hyperbolic ring, since the whole of the dissipative quantum dynamics is realized by choosing only one representation of the field commutation relations.
Entanglement entropy operators for the subsystem of interest and the environment, are constructed as a tool for study the entanglement generated from the dissipation. The holographic perspectives of our results are discussed.\\

\noindent KEYWORDS: hyperbolic symmetries; quantum dissipation; entanglement entropy; holography.

\section{Antecedents, motivations, and results}
At microscopic level, the existence of quantum entanglement has been demonstrated in systems involving photons, electrons, ions, etc;  however, although at macroscopic level the entanglement is sensitive to environmental perturbations, such a phenomenon has been inferred
by several experiments involving macroscopic-scale objects; for example in \cite{nature}, two massive micromechanical oscillators are coupled to an electromagnetic cavity, and the existence of entanglement is inflicted from the measurements of mechanical fluctuations, and analysis of the microwaves coming from the cavity. Furthermore,  
it has been shown experimentally that
dissipation generates entanglement between two macroscopic objects \cite{cirac}; dissipative dynamics is intrinsically stable under perturbations, stabilizing the entanglement, in contrast to other schemes based on coherent evolution, which is always susceptible to 
undergo decoherence. The entanglement generated in this experiment corresponds to an EPR type one, and thus relevant in areas such as
quantum information, cryptography, and others.

Dissipative quantum field theories are of interest in high energy physics; for example, finite temperature techniques are necessary for estimating the transport coefficients of the quark-gluon plasma, and the holographic techniques have allowed to obtain results on the viscosity, in a supersymmetric version of the Yang-Mills plasma
\cite{policastro}. Furthermore, the Brownian motion of a heavy quark has been studied from different points of view, from holographic black holes \cite{npb}, Schwinger-Keldysh holographic scheme \cite{boer}, and thermal field dynamics \cite{thft}; the results suggest that the dissipative e\-ffects generated by the emission of gluonic quanta, are relevant in the quark dynamics even at zero temperature, in consistency with the conclusions obtained later in \cite{baner}. 

Dynamical entanglement generated by dissipation has been studied within the context of the ADS/CFT correspondence \cite{botta}; since holographic correspondence relates geometrical quantities with information coming from quantum field theory, in this reference the Ryu-Takayanagi formulae for the area of a minimal surface in a holographic geometry, is used for studying the entanglement between the degrees of freedom of dissipative conformal field theories; the holographic correspondence suggests then a teleportation protocol that interchange quantum information between the two conformal field theories on the boundaries of the dual geometry; such a protocol is similar to the mechanism that makes a wormhole traversable \cite{gao,malda}.

On the other hand, low dimensional condensed matter systems exhibit dissipation, and a holographic interpretation must be explored within the gravity/condensed-matter duality along the lines described above; for example, in the dynamics of a point-like impurity in a Bose-Einstein condensate at zero temperature, the dissipation is generated by the emission of elementary excitations, such as phonons \cite{fonon}; a similar mechanism works in the dissipation of quantum turbulence in superfluid $^{4}He$ \cite{4H}. Furthermore, in electron systems the intrinsic decoherence (dissipation) can be caused by
electron-phonon, electron-electron, and magnetic impurity interactions; a set of experiments with gold wires have been designed for understanding  the functional dependence of the phase coherence time on the temperature \cite{mesos}. Additionally the systems with strong interlayer correlations 
exhibit drag transresistence and transconductance even at zero temperature \cite{resis}.

Usually, it is assumed that the holographic duality represents a tool that allows us to use general relativity for describing other areas of physics, typically superconductivity by using black hole thermodynamics; moreover, one of the main applications of the correspondence it is to gain insight into strongly coupled gauge theory, for example, for providing an alternative geometrical description of the confinement (see \cite{horowitz}, for a pedagogical review). Conversely, the holographic duality allows us, in principle, to use gauge theories and condensed matter systems for gaining insight into quantum gravity. However, beyond the holographic duality/enthusiastic community,  critical points of view for demystifying the holographic mystique have been also raised \cite{dv}, with the purpose of ascertaining the true status and to discuss the bases on which the holographic predictions have been obtained. Along the same critical lines, R. Penrose has posed severe criticisms on holographic principles in his books \cite{rp1}, and \cite{rp2}.
In the present work we are no committed with one side of the enthusiastic/critical balance in our holographic perspectives; first, our reformulation of quantum dissipation, is independent on possible holographic applications; second, with a noncanonical quantization scheme at hand, we follow in parallel previous applications that use conventional quantization schemes in holographic scenarios;  hence, we pose the bases for replying a simple question: whether (or not) the holographic realizations are independent on the quantization scheme.

In this paper we shall follow closely the developments of the reference \cite{botta}, which in its turn, followed closely the scheme developed in \cite{cele} for the quantum dissipation; in the first part of that reference, the classical and quantum formulations for dissipative field theory 
are constructed using a conventional scheme, {\it i.e.} based on the conventional complex plane; in particular, a {\it canonical quantization} is formulated along the universal prescriptions  of the quantum field theory formalism. However, in the present paper, by considering the same Lagrangian formulation used in that reference, we realize that a hyperbolic complex structure underlies the dissipative dynamics, reveling thus the existence of an internal noncompact symmetry that drives the dynamics. Hence, the subsequent quantum formulation will be committed with that hyperbolic complex structure; hence, a bifurcation emerges respect to the formulation based on the elliptic (conventional) complex plane, leading to a new quantum field theory for the dissipation. Although it is well understood that the dissipation  breaks explicitly the background Lorentz symmetry, the present treatment shows that a {\it Lorentzian} internal group related with hyperbolic rotations in field space, is present indistinctly whether the dissipation is switched on or switched off. Thus, the hyperbolic scheme provides an appropriate and alternative description of the classical and quantum dissipative dynamics, and in principle a broadly applicable theoretical technique, for example for studying and reformulating the different dissipative systems described previously in both high energy and condensed matter scenarios. 
The main virtues of the scheme developed here are, for example, to cure the pathologies associated with existence of many unitarity inequivalent representations, and to cure the non-preservation of the field commutators due to the presence of damping factors, which have been persistent problems in the traditional treatments on dissipative dynamics; 
additionally within the present scheme the field commutators  depend in a novel way on the dissipative parameter  and on the background dimension, since they do not correspond simply to Dirac delta distributions; similar results are obtained for physical observables and operators. As an explicit example, the case of $1+1$ dissipative quantum field theory, of wide interest in condensed matter systems of low dimensionality,  will be developed in exact form, and it will show profound differences with respect to the conventional treatments.

It is worthwile to mention that as the manuscript \cite{cele}, is the basic reference on dissipative quantum field theory (DQFT), whose formulation is constructed on the quantum dynamics for a damped harmonic oscillator; thus, the fundamental aspects of DQFT such as the dynamical symmetry algebra, field commutation relations, construction of observables, etc, are inherited from the physics of a harmonic oscillator. In the present scheme we can not, in fact,
 to follow the quantization of a harmonic oscillator, since by identifying the symmetries of the dissipative system, and the corresponding hyperbolic complex plane, we shall be strongly constrained by the algebraic structure of the hyperbolic ring.
For achieving this goal, in the next section we introduce the hyperbolic formalism, by describing the hyperbolic complex plane, which is little known in the physics literature.

Previous treatments using hyperbolic complex plane include for example
the unification of diverse scalar fields of cosmological and particle physics interest
\cite{cuaglia}; in particular an inflaton-phantom cosmology has been established without invoking concepts such as field theory {\it Euclideanization}, and the use of noncanonical forms for the Lagrangian terms. As opposed to traditional treatments on the phantom fields dynamics, bounded potentials are constructed along the universal prescriptions, and the spontaneous symmetry breaking physics can be studied in a conventional way in cosmological scenarios.
Additionally
the use of  the hyperbolic formalism has allowed the formulation of hypercomplex gauge field theories \cite{Oscar}; the hypercomplex electrodynamics includes hyperbolic counterparts of gauge fields, Higgs fields, and Nambu-Goldstone bosons. The physics of topological defects associated with continuous symmetries  turns out to be radically different from that established for conventional gauge field theories. Furthermore, the effective value for the coupling constants in hypercomplex electrodynamics are comparable with those generated by quantum fluctuations \cite{defor}.

In the section \ref{lag}, we start with a Lagrangian that consists of doubling the degrees of freedom of the system of interest by defining a  
copy thermal field representing the environment, in consistency with the thermo field dynamics scheme. Since the identical copy field that represents the environment evolves in the inverse time direction, the corresponding Lagrangian terms appear in this scheme with an unconventional minus sign \cite{santana}; this Lagrangian structure represents the key for the realization of an underlying hyperbolic structure in the dissipative dynamics, and our starting point for the new classical and quantum formulation for dissipative field theories.
In Section \ref{ovd}  it is shown that the quantization is noncanonical, in the sense that the field commutators do not satisfy the usual canonical commutation relations found in the usual approaches; instead of this, the new commutation relations will reveal an underlying non-commutative field theory, which will emerge naturally as consequence of the algebraic structure of the hyperbolic ring, as opposed to other approaches; for example, in {\cite{pra}} the quantum effects on a pair of (damped) harmonic oscillators embedded in the Moyal plane are analyzed, and the results show that the noncommutativity acts as a source of dissipation; in this approach the Moyal plane is introduced by hand, since it is assumed as the ambient space from the beginning. Furthermore, in \cite{prd}, it is shown that the noncommutative spectral geometry is related to dissipation through a common fundamental ingredient, namely,  the concept of doubling of the algebra; thus, with the appearance of an {\it interference phase}, the dissipation plays the key role in the quantization, according to the t'Hooft's conjecture.
 In Section \ref{modular1} the Hamiltonian operator and the corresponding evolution operator are considered; 
the quantum field theory constructed evolves unitarily, and the dissipative effects are developed by the system of interest under the Hamiltonian dynamics of the entire system. Furthermore,
since a conserved charge is obtained from such a hyperbolic continuous symmetry, a chemical potential conjugate to that charge is identified naturally; thus, one of the important results of the present treatment is the identification of a grand partition function with the duplication of degrees of freedom in the double field theory formalism; the full partition function is discussed in the subsection \ref{chemical}. The vacuum state and the vacuum expectation values for quantum observables such as the Hamiltonian and the charge are discussed in Section \ref{coherent}. In section \ref{states}  the evolved vacuum state is constructed as an entangled two-mode state; in particular asymptotic entangled states are constructed. Also in this section, the traditional pathologies associated with a conventional quantum field theory approach for the dissipation are cured, establishing one important result from the non-conventional approach developed here.
 In section \ref{entropy}, time dependent entanglement 
entropy operators and their expectation values at the evolved vacuum  are derived in terms of solutions of certain integral equations. Possible holographic scenarios are discussed at the end, as prospects for the future. Finally, we introduce some concluding remarks and perspectives in section \ref{conlusion}.

\section{The formalism: the hyperbolic ring}

The real number system can be extended in three forms, $x+\iota y$, where $\iota$ can be any of the three possible complex units, namely, the parabolic one with $\iota^{2}=0$, the elliptic one with $\iota^{2}=-1$ (which corresponds to the usual complex unit), and the hyperbolic one with $\iota^{2}=1$, which will play a fundamental role in the present treatment. Further commutative and noncommutative extensions that involve more than one complex unit can be constructed \cite{kisil,ulrich, pood, sob}. The hyperbolic complex unit will be termed as $j$ from this point, and in the following we describe in detail the algebraic structure of the corresponding {\it hyperbolic complex plane}. The closure respect to the addition and multiplication is fully similar to the usual complexes; however the hyperbolic squared modulus is not positive definite,
\begin{equation}
     z\bar{z}= x^{2}-y^{2}, \label{modulus}
\end{equation}
where $z=x+jy$, $\bar{z}=x-jy$, with $\bar{j}=-j$, and $j^{2}=1$.
The modulus (\ref{modulus}) is invariant under the hyperbolic rotation $z\rightarrow e^{j\chi}z$, with $e^{j\chi}=\cosh \chi + j\sinh \chi$, where $\chi$ is a noncompact parameter $\chi \in R$. These continuous rotations correspond to the connected component
of the Lie group $SO(1,1)$ that contains the identity element of the group $I$ with $\chi=0$.

The multiplicative inverse is defined as
\begin{eqnarray}
     z^{-1} = \frac{\bar{z}}{z\bar{z}} = \frac{x-jy}{x^{2}-y^{2}}, \label{inverse}
\end{eqnarray}
provided that $x\neq\pm y$. The cases $x=\pm y$ define two {\it isotropic} lines
 that separate the quadrants of the hyperbolic plane, and define also an idempotent basis.

As opposed to the usual complexes with the idempotents 0 and 1, the ring ${\cal H}$ possesses in addition nontrivial idempotents,
\begin{eqnarray}
     J^{+} \! & = & \! \frac{1}{2} (1+j), \quad (J^{+})^{n} = J^{+}; \nonumber \\
     J^{-} \! & = & \! \frac{1}{2} (1-j), \quad (J^{-})^{n} = J^{-}, \quad n=1,2,3 \ldots ;
     \label{idem}
\end{eqnarray}
furthermore, it is easy to prove that they annihilate to each other,
\begin{equation}
     J^{+} J^{-}=0, \label{ortho}
\end{equation}
and hence the ring ${\cal H}$ is not an integral domain \cite{pood}; moreover, the hyperbolic unit $j$ is absorbed by these idempotents
\begin{equation}
     jJ^{+}=J^{+}, \qquad jJ^{-}=-J^{-}; \label{absorb}
\end{equation}
note the presence of a minus sign in the second identity. The idempotents work like {\it projectors} in the ring ${\cal H}$; for an arbitrary hyperbolic number, $z=x+jy$, we have that
\begin{equation}
     J^{+}z = (x+y)J^{+}, \quad J^{-}z=(x-y)J^{-}, \quad z=J^{+}z+J^{-}z; \label{projector}
\end{equation}
where the identities (\ref{absorb}) have been taken into the account. In the present scheme, the classical fields, quantum operators, state vectors will be described in terms of these projectors.
In particular, the hyperbolic exponential can be expressed in terms of its projections
\begin{equation}
     e^{j\chi} = e^{\chi}J^{+} + e^{-\chi} J^{-}, \quad \chi \in R.
\label{projector1}
\end{equation}
This implies also that,
\begin{eqnarray}
e^{J^+\chi} = J^{+}e^{\chi}+J^{-}, \quad e^{J^-\chi} = J^{-}e^{\chi}+J^{+}; 
\label{projector2}
\end{eqnarray}
quantities of this form, will have norm positive definite, since
$|e^{J^+\chi}|^{2}= |e^{J^+\chi}|^{2}\equiv e^{J^+\chi}e^{J^{-}\chi}=e^{\chi}$, and $\sqrt{|e^{J^+\chi}|^{2}}=e^{\frac{1}{2}\chi}$; 
thus, the expressions in Eq. (\ref{projector2}) are basically inverse to each other;
certain operators and their expectation values will take this form. The series expansion will be required, and it can be obtained from the expressions on the right hand side, considering first the usual expansion for the real function $e^{\chi}$, and then 
\begin{eqnarray}
e^{J^{+}\chi}=J^{+}\sum_{n=0}^{\infty}\frac{\chi^{n}}{n!}+J^{-}=J^{+}+J^{-}+J^{+} \sum_{n=1}^{\infty}\frac{\chi^{n}}{n!}=1 +J^{+} \sum_{n=1}^{\infty}\frac{\chi^{n}}{n!},
\label{serie}
\end{eqnarray}
and similarly for the second case, which can be obtained directly by complex conjugation from the above expression; since the infinite series
in the above equation is constructed for $n\geq 1$, it is well defined when 
$\chi$ corresponds to an operator; as we shall see later,
the detailed description of entangled states will require these expansions.

At this point, it is difficult to understand the relation between this formalism and the dissipation phenomena; however, we shall see in the next section that the group $SO(1,1)$
can be reveled as an underlying internal symmetry in a Lagrangian description of the dissipation, even though the background Lorentz symmetry is explicitly broken. Moreover, the projectors of the idempotent basis will determine a noncanonical quantization scheme for the dissipative dynamics.

\section{The dissipative Lagrangian}
\label{lag}
The starting point is the following Lagrangian constructed on a $d$-dimensional Minkowskian background,
\begin{equation}
     L(\Phi,\Psi) = \frac{1}{2} \int dx^{d} [(\partial_{\mu} \Phi )^{2} - (\partial_{\mu}\Psi )^{2} + \gamma (\Psi\dot{\Phi} - \Phi\dot{\Psi})],
     \label{l}
\end{equation}
which corresponds to Eq.(11) in \cite{botta}; $\gamma$ is the damping coefficient, $(\Phi,\Psi)$ are real fields, and a dot denotes time derivative. Now, we define the following hyperbolic field
\begin{equation}
     \Omega = \Phi + j\Psi ; \label{omega}
\end{equation}
and thus we construct the following real objects that are invariant under global hyperbolic rotations, and under ${\cal PT}$-like transformation
in field space,
\begin{eqnarray}
     \partial_{\mu}\Omega \cdot \partial^{\mu}\overline{\Omega} &=& (\partial_{\mu}\Phi )^{2} - (\partial_{\mu}\Psi)^{2}, \quad SO(1,d-1), \quad SO(1,1); \label{lorentz1}\\
     \frac{1}{2} [j\Omega\dot{\overline{\Omega}} + c.c.] &= &\Psi\dot{\Phi} - \Phi\dot{\Psi},\quad SO(1,1); 
     \label{lorentz2}
\end{eqnarray}
which reproduce exactly the terms appearing in the Lagrangian (\ref{l}); we have indicated the symmetry groups for each term, and hence, the Lorentz
symmetry is explicitly broken in the second term, since the boost transformations are not preserved, and the dissipation defines an "arrow of time", with the thermal bath frame as the preferred frame.
Therefore, the Lagrangian can be rewritten explicitly as a $SO(1,1)$-invariant,
\begin{equation}
     L(\Omega,\overline{\Omega}) = \frac{1}{2} \int dx^{d} [\partial_{\mu}\Omega\cdot\partial^{\mu}\overline{\Omega} + \frac{\gamma}{2} (j\Omega\dot{\overline{\Omega}} + c.c.)], \quad {\rm modulo} \quad \Omega\rightarrow e^{j\chi}\Omega,\quad {\rm and} \quad \Omega\rightarrow -\Omega.
     \label{l1}
\end{equation}
If the dissipative interaction between the fields is effectively interpreted as a sort of quantum quen\-ching \cite{quen}, then the internal symmetry $SO(1,1)$ will survive the suddenly switched on, and the consequential energy exchange.

Now we re-construct the original action in \cite{botta}, using a redefinition of fields proposed in  Eq. (10) of that reference,
\begin{equation}
     \Phi = \frac{\phi +\psi}{\sqrt{2}}, \quad \Psi = \frac{\phi -\psi}{\sqrt{2}}; \label{l2}
\end{equation}
and then our hyperbolic field (\ref{omega}) reads
\begin{equation}
     \frac{\Omega}{\sqrt{2}} = J^{+}\phi + J^{-}\psi; \label{l3}
\end{equation}
therefore, $(\Phi,\Psi)$ correspond to the components of $\Omega$ in the standard basis, and $(\phi,\psi)$ correspond to the components in the idempotent basis. However, the transformation (\ref{l2}) can be reproduced using the projectors (\ref{projector}) on the field (\ref{omega}), $J^{+}\Omega = (\Phi +\Psi)J^{+}$, and $J^{-}\Omega = (\Phi -\Psi)J^{-}$, which correspond essentially to the definition of the fields $(\phi,\psi)$ in Eq. (\ref{l2}).

Therefore, the substitution of the expression (\ref{l3}) into the Lagrangian (\ref{l}) leads to,
\begin{eqnarray}
     L(\phi,\psi) \! & = & \! \int dx^{d} \{ J^{+}[\partial_{\mu}\phi\cdot\partial^{\mu}\psi + \frac{\gamma}{2} (\phi\dot{\psi} - \psi\dot{\phi})] + J^{-} [\partial_{\mu}\phi\cdot\partial^{\mu}\psi + \frac{\gamma}{2} (\phi\dot{\psi} - \psi\dot{\phi})]\}, \label{l4} \\
     \! & = & \! \int dx^{d} [\partial_{\mu}\phi\cdot \partial^{\mu}\psi + \frac{\gamma}{2}(\phi\dot{\psi}-\psi\dot{\phi})]; \label{l5}
\end{eqnarray}
Eq. (\ref{l5}) is exactly the Lagrangian (4) in Ref. \cite{botta}, which describes the dissipative behavior through the interaction between the field $\phi$ and its identical copy $\psi$. Hence, the equations of motion read,
\begin{equation}
     (\partial_{t}^{2}-\nabla^{2})\phi + \gamma\partial_{t}\phi=0, \quad (\partial_{t}^{2}-\nabla^{2})\psi - \gamma\partial_{t}\psi=0.
     \label{ecm}
\end{equation}

\subsection{The equation of motion and its spectral decomposition}
\label{eqm}
Once we have reproduced the conventional description of the dissipation phenomenon used in \cite{botta}, we return to the action (\ref{l1}), in which the dynamical variables correspond to the pair $(\Omega ,\bar{\Omega})$; the variation respect to $\bar{\Omega}$ leads to the equation
\begin{equation}
     \partial_{\mu}\partial^{\mu}\cdot\Omega + j\gamma\partial_{t}\Omega =0, \label{euler}
\end{equation}
similarly the variation with respect to $\Omega$ leads to the complex conjugate version of the above equation. Now, the substitution of the expression (\ref{l3}) into the equation (\ref{euler}) yields the {\it spectral decomposition} of the equations of motion,
\begin{equation}
     J^{+} [\partial_{\mu}\partial^{\mu}\phi + \gamma\dot{\phi}] + J^{-} [\partial_{\mu}\partial^{\mu}\psi -\gamma\dot{\psi}] =0;
     \label{eulerspectro}
\end{equation}
where the properties (\ref{absorb}) have been considered; hence the equations of motion (\ref{ecm}) are replicated exactly from the Eq. (\ref{eulerspectro}).

\subsection{The solution}
Now we can propose a hyperbolic phase as a solution to the equation of motion, that is decomposable in the base $(J^+, J^-)$ according to the Eq. (\ref{projector1}); furthermore, since the equation of motion admits the spectral decomposition (\ref{eulerspectro}) we propose an arbitrary combination of solutions in the $(J^{+},J^{-})$ basis
\begin{eqnarray}
\Omega (\vec{x},t) = (aJ^{+}) e^{-\frac {\gamma t}{2}} e^{w_{_{k}}t-\vec{k}\cdot \vec{x}} + (\bar{b}J^{-}) e^{\frac {\gamma t}{2}} e^{-(w_{_{k}}t-\vec{k}\cdot \vec{x})} ,
\label{spectro1}
\end{eqnarray} 
where $(a,b)$ are arbitrary coefficients and the spectral parameters $(w,\vec{k})$ are real-valued; the substitution into the Eq. (\ref{euler}) leads to
\begin{eqnarray}
(w_{_{k}}^{2}-\vec{k}\cdot \vec{k} - {\frac {\gamma ^2 }{4}}) \Omega = 0, \quad \rightarrow \quad w_{_{k}}= \pm \sqrt{\vec{k}\cdot \vec{k} + {\frac {\gamma ^2 }{4}}}; 
\label{spectro2}
\end{eqnarray} 
where the properties (\ref{absorb}) have been used, and they play a key role in factoring out the solution $\Omega$, and thus, both components $J^{+}$, and $J^{-}$ have the same {\it dispersion} relation.
Therefore, the IR limits are well defined, 
\begin{eqnarray}
\lim_{\vec{k} \rightarrow 0} w_ {_k}  = \pm \frac {\gamma}{2}. \label {if2}
\end{eqnarray} 
However, within the scheme employed in \cite{botta}, the solution is described in terms of quasi-normal frecuencies, with a different dispersion relation,
\begin{equation}
   \omega_{k} = \pm\sqrt{\vec{k}\cdot\vec{k}-\frac{\gamma^{2}}{4}}; \quad
   \quad {\rm conventional} \quad {\rm scheme;}
    \label{cutoff}
\end{equation}
and thus an IR cut off is imposed by $\gamma$, since the $\lim_{\vec{k}\rightarrow 0}\omega_{k}$ is not defined as a real frequency. We need explain this issue in more detail, since we have the same equations of motions in both schemes but with different dispersion relations. Considering that the real exponentials in Eq. (\ref{spectro1}) are solutions to the equations of motion, the analytically continued solutions can be obtained  through the transformations $\omega_{k}\rightarrow i\omega_{k}$, and $k\rightarrow i k$, within the standard scheme, and through $\omega_{k}\rightarrow j\omega_{k}$, and $k\rightarrow j k$ within the case at hand, which lead to the purely hyperbolic expression (\ref{spectro1}). In both cases the complexification does not change the relative sign between $\omega_{k}^2$, and $k^2$ in the dispersion relations, but the relative sign of the dissipative term $\gamma^2$ is different in each case, since in the former $i^2=-1$, and in the later $j^2=1$. Thus, for the same equations of motion, the corresponding dispersion relation depends on which complex plane the solutions are analytically continued.
If one attempts to solve the hyperbolic ring based differential equation (\ref{euler}), by using the usual complex phase, then the product $ij$ will appear, which has not been defined within the ring ${\cal H}$; at this point one may extend the ring for including the usual complex unit $i$, and then to proceed  with the quantization prescription (see \cite{alex}). In fact, when the solution for equations of motion considered previously is continued into such an extended ring that includes both complex units $(i,j)$, then the corresponding dispersion relation turns out to be that in Eq. (\ref{cutoff}) \cite{rcf}.
Thus, we prefer, in the approach at hand, to work strictly within the ring ${\cal H}$, and hence only the complex unit $j$ is available. The quantization based on quantum operators will be committed with this hyperbolic complex unit coming from the classical formulation of the dissipation, leading to a new quantum description of such a phenomenon. As a complementary approach, in the section \ref{path}  we develop a path integral formulation, which requires additionally the use of the conventional complex unit $i$. 

\section{Operator-valued distributions}
\label{ovd}
If the spectral decomposition of the field operator is constructed from the  solution (\ref{spectro1}), considering the full range k  $\in(-\infty,+\infty)$, taking into account the usual Dirac delta functions resulting from the commutators between the annihilaton and creation operators,, then the field operator commutators will undergo UV divergences; therefore we have to re-consider the decomposition range and the distribution representing the Dirac delta on the hyperbolic ring.

We start with the following sequence of functions that defines the Dirac delta an a limit by the right side
\begin{equation}\label{delta1}
\delta_{n}(x)=\left\{
\begin{array}{ll}
0,& \quad x< 0;\\
ne^{-nx},& \quad x>0, \quad n = 1,2,3\ldots;
\end{array}
\right.
\end{equation}
this delta sequence satisfies 
\begin{eqnarray}
\lim_{n\rightarrow +{\infty}} \int^{+\infty}_{-\infty} \delta_{n} (x) f(x) dx = \lim_{n\rightarrow +{\infty}} \int^{+\infty}_{0} \delta_{n}^{+} (x) f(x) dx = f(+0), 
\label{delta2} 
\end{eqnarray}
where the singularity is at $+0$, the positive side of the origin, and we have defined $\delta_{n}^{+} (x) = \delta_{n} (x>0)$.
In order to define a spectral decomposition of the fields, we consider first that the full field defined in Eq. (\ref{l3}), $\frac{\Omega}{\sqrt{2}} = J^{+}\phi + J^{-}\psi$,  represents the complete system in the basis $(J^+,J^-)$, including the subsystem A ($\phi$) , and the environment ($\psi$);  thus, the full range of the spectral parameters  $\vec{k}$ for the complete system, must be splitted into sub-ranges; such a splitting depends on the physical properties of the subsystems, and we propose here an {\it equipartition} of the momenta according to
\begin{eqnarray}
(-\infty, 0) \longleftrightarrow J^{-}, \quad {\rm} \quad (0,+\infty) \longleftrightarrow J^{+};
\label{delta3}
\end{eqnarray}
hence the spectral decomposition of the field operator constructed from the solution (\ref{spectro1}) reads
\begin{eqnarray}
\hat{\Omega} (\vec {x} , t) = N [J^{+} \int\limits ^{+\infty}_{0} \hat{a} ( \vec {k}) e^{-\frac {\gamma t}{2}} e^{w_{_{k}}t-\vec{k}\cdot \vec{x}} d \vec{k} +  J^{-} \int\limits ^{0}_{-\infty} \hat{b}^{\dagger} ( \vec {k}) e^{\frac {\gamma t}{2}}  e^{-(w_{_{k}}t-\vec{k}\cdot \vec{x})} d \vec{k}]
\label {operator1} 
\end{eqnarray} 
where $N$ is a normalization factor that will be fixed  later. Now we construct the following field commutator,
\begin{eqnarray}
[\hat{\Omega} (\vec{x},t), \hat{\Omega}^{\dagger} (\vec{x'},t)] 
= N^{2} \int^{+\infty}_{0} d\vec{k} \int^{+\infty}_{0} d\vec{k'}{e^{(w_{k}-w_{k'})t}}\Big[J^{+} [\hat{a}(\vec{k}),\hat{b}(\vec{-k'})] e^{\vec{-k'}\cdot \vec{x'}-\vec{k}\cdot \vec{x}} \nonumber \\ +  J^{-}[\hat{b}^\dag(\vec{-k'}),\hat{a}^\dag(\vec{k})]e^{\vec{-k'}\cdot \vec{x}-\vec{k}\cdot \vec{x'}}\Big],
\label{cc2} 
\end{eqnarray}
 terms of the form $e^{-\frac {\gamma t}{2}}(J^{+}J^{-})[\hat{a},\hat{a}^{\dagger}]$, and 
$e^{\frac {\gamma t}{2}}(J^{+}J^{-})[\hat{b},\hat{b}^{\dagger}]$ vanish trivially due to
the property (\ref{ortho});  therefore, an algebraic feature of the present scheme is to avoid the non-preservation of the commutation relations under time evolution due to the presence of damping terms, a persistent problem within  conventional dissipative-QFT schemes. This fact defines a bifurcation point between the standard scheme and the scheme at hand; the former is based on the damped harmonic oscillator quantization, for which the commutators $[\hat{a},\hat{a}^{\dagger}]$, and $[\hat{b},\hat{b}^{\dagger}]$, are essential, and thus the presence of damping factor is irremovable. Due to this fact, and  although the formulations of the dissipative systems have been widely studied, all treatments have suffered historically from this problem; in fact, in the traditional schemes there exists a 
physical mechanism for preserving the canonical structure, the fluctuating forces \cite{salerno}. For an alternative description of dissipative quantum mechanics, and an updating on the different formulations to the date, describing those long-lived problems, see for example \cite{cupa}.

As we shall see, all field commutators within the present scheme will show this feature, the absence of damping factors; the additional dependence on the damping parameter $\gamma$ through the frequencies $w_{k}$, and $w_{k'}$, will be eliminated fully once we have chosen a distribution for the commutator $[\hat{a}, \hat{b}]$. Furthermore, 
 the field commutator (\ref{cc2}) vanishes within a standard quantization scheme,  since there, one usually fixes the commutator $[\hat{a}(\vec{k}),\hat{b}(\vec{-k'})] =0$; however,  within the present scheme, all field commutators depend basically on this commutator, and hence it must be maintained as nonvanishing.
Now we assume the following commutation rule for the creation and annihilation operators,
\begin{eqnarray}
[\hat{a}(\vec{k}),\hat{b}(\vec{k'})]  =  \rho \delta^{+} (\vec{k}-\vec{k'}), \quad
[\hat{b}^{\dagger}(\vec{k'}),\hat{a}^{\dagger}(\vec{k})]  =  \bar{\rho} \delta^{+}  (\vec{k}-\vec{k'}), \quad \rho = \rho_{1} + j \rho_{2},\label{fc2}
\end{eqnarray}
where the delta Dirac is defined by means of the sequence established in Eq.  (\ref{delta1}), and $\rho$ is an arbitrary element in the ring that in general depends on ($\vec{k},\vec{k'}$).

Therefore the commutator (\ref{cc2}) reduces to
\begin{eqnarray}
[\hat{\Omega} (\vec{x},t), \hat{\Omega}^{\dagger} (\vec{x'},t)]
=N^{2}(\rho_{1}+\rho_{2})\int^{0}_{-\infty} [J^{+}e^{\vec{k'}\cdot (\vec{x'}-\vec{x})}+J^{-}e^{-\vec{k'}\cdot (\vec{x'}-\vec{x})} ]d\vec{k'};
\label{owitho}
\end{eqnarray}
which in general diverges. In order to address this issue, let us consider certain restrictions which allow us to determine convergence criteria. If $x'_{i}-x_{i} > 0$, we observe that 
\begin{eqnarray}
\int^{0}_{-\infty} e^{\vec{k}\cdot(\vec{x'}-\vec{x})} d\vec{k} =\prod_{i} {\frac{1}{x'_{i}-x_{i} }} , \quad \int^{0}_{-\alpha} e^{\vec{k}\cdot(\vec{x'}-\vec{x})}d\vec{k}= \infty ;
\label{diver1}
\end{eqnarray}
thus, using the property (\ref{ortho}) we can eliminate the divergent $J^{-}$-term, and to obtain a $J^{+}$-projected distribution for the field commutator, 
\begin{eqnarray}
J^{+}[\hat{\Omega} (\vec{x},t),\hat{\Omega}^{\dagger} (\vec{x'},t)] = N^{2} (\rho_{1}+\rho_{2})\prod_{i} {\frac{J^{+}}{x'_{i}-x_{i} }},  \quad x'_{i}-x_{i} > 0;
\label{diver2}
\end{eqnarray}
and conversely we can construct the $J^{-}$-complement,
\begin{eqnarray}
J^{-}[\hat{\Omega} (\vec{x},t),\hat{\Omega}^{\dagger} (\vec{x'},t)] = N^{2} (\rho_{1}+\rho_{2})(-1)^l\prod_{i} {\frac{J^{-}}{x'_{i}-x_{i} }} , \quad x'_{i}-x_{i}< 0;
\label{diver3}
\end{eqnarray}
in this expression $l$ corresponds to the number of spatial dimensions;
in terms of the {\it sign} function $sgn(x)$ that is $+1$ for $x>0$, and $-1$ for $x<0$, such a commutator can be written as 
\begin{eqnarray}
[\hat{\Omega} (\vec{x},t),\hat{\Omega}^{\dagger} (\vec{x'},t)] = N^{2} (\rho_{1}+\rho_{2})
\Big [{J^{+}}\prod_{j}\frac{sgn (x'_{i}-x_{i} ) +1}{2} +(-1)^{l+1} {J^{-}}\prod_{j}\frac{sgn(x'_{i}-x_{i} )-1}{2}\Big]\prod_{i} {\frac{1}{x'_{i}-x_{i} }};
\label{sgn1}
\end{eqnarray}
Moreover, such a commutator has the following physically admissible limits
\begin{eqnarray}
\lim_{\vec{x'}- \vec{x} \rightarrow 0} [\hat{\Omega} (\vec{x},t),\hat{\Omega}^{\dagger} (\vec{x'},t)] = \infty, \quad
\lim_{\vec{x'}- \vec{x} \rightarrow \infty} [\hat{\Omega} (\vec{x},t),\hat{\Omega}^{\dagger} (\vec{x'},t)] = 0.
\label{sgn2}
\end{eqnarray}
Now, from Eq. (\ref{l1}) we can determine the canonical conjugate momenta of the fields $(\Omega ,\bar{\Omega})$,
\begin{equation}
     \Pi_{\Omega} \equiv \frac{\partial L}{\partial\dot{\Omega}} = \dot{\bar{\Omega}} - \frac{\gamma}{2} j\bar{\Omega} =\Pi_{\Phi} +j\Pi_{\Psi}, \quad \Pi_{\bar{\Omega}} = \Pi_{\Phi}- j\Pi_{\Psi}; \label{momenta}
\end{equation}
which have been written in terms of the momenta constructed from the Lagrangian (\ref{l}) for the fields $(\Phi ,\Psi)$;
therefore, the conjugate momentum operator can be constructed from this classical expression, 
 \begin{eqnarray}
\hat{\Pi} _{\Omega}(\vec{x},t) = N[J^{-}\int\limits ^{+\infty}_{0}{w_{k}}\hat {a}^{\dagger} (\vec {k}) e^{-\frac {\gamma t}{2}} e^{w_{_{k}}t-\vec{k}\cdot \vec{x}}d \vec{k} - J^{+}\int\limits ^{0}_{-\infty}{w_{k}} \hat {b} (\vec {k}) e^{\frac {\gamma t}{2}} e^{-(w_{_{k}}t-\vec{k}\cdot \vec{x})} d \vec{k}] .
\label {operator11} 
 \end{eqnarray}
With these expressions at hand, we are allowed  to construct the following equal-time field commutator 
\begin{eqnarray}
[\hat{\Pi}_{_\Omega} (\vec{x},t), \hat{\Pi}^{\dagger}_{_\Omega} (\vec{x'},t)]=  N^{2} \int^{+\infty}_{0} d\vec{k} \int^{0}_{-\infty} d\vec{k'}w_{k} w_{k'} e^{(w_{k}-w_{k'})t}\Big[J^{+} [\hat{a}(\vec{k}),\hat{b}(\vec{k'})] e^{\vec{k'}\cdot \vec{x}-\vec{k}\cdot \vec{x'}} \nonumber \\+ J^{-}[\hat{b}^\dag(\vec{k'}),\hat{a}^\dag(\vec{k})]e^{\vec{k'}\cdot \vec{x'}-\vec{k}\cdot \vec{x}}\Big]
\nonumber \\
 =  N^{2}(\rho_{1}+\rho_{2})\int^{0}_{-\infty}w^{2}_{k'} [J^{+}e^{-\vec{k'}\cdot (\vec{x'}-\vec{x})}+J^{-}e^{\vec{k'}\cdot (\vec{x'}-\vec{x})} ]d\vec{k'};
\label{piwithpi}
\end{eqnarray} 
which must be compared directly with the first commutator (\ref{owitho}); note in particular the presence of a quadratic factor in the frequency, which is depending on $\vec{k}$
according to the Eq. (\ref{spectro2}).  The indefinite integral $\int e^{au}u^2du= \frac{e^{au}(a^2u^2-2au+2)}{a^3}$ allows us to solve the integral proportional to $\vec{k}\cdot\vec{k}$; the integral proportional to $\gamma^2$ can be solved exactly along the procedure used for obtaining (\ref{owitho}). Therefore, the commutator reads
\begin{eqnarray}
[\hat{\Pi}_{_\Omega} (\vec{x},t), \hat{\Pi}^{\dagger}_{_\Omega} (\vec{x'},t)]=   N^{2}(\rho_{1}+\rho_{2})
\Big [\prod_{j}\frac{sgn (\vec{x'}- \vec{x})_{j} +1}{2}{J^{-}} +(-1)^{l+1} \prod_{j}\frac{sgn(\vec{x'}- \vec{x})_{j}-1}{2} {J^{+}}\Big]\cdot \nonumber\\
\prod_{i} {\frac{1}{(\vec{x'}- \vec{x})_{i}}}\Big[ \sum_{j}\frac{1}{(\vec{x'}- \vec{x})^{2}_{j}} +\frac{\gamma^2}{4}\Big],
\label{piwithpi2}
\end{eqnarray} 
which satisfied the same limits described in Eq. (\ref{sgn2}); note that this commutator depends explicitly on the $\gamma$-parameter, as opposed to the first commutator (\ref{sgn1}).

Along the same lines, we construct the fundamental
equal-time commutator,
\begin{eqnarray}
[\hat{\Omega} (\vec{x},t), \hat{\Pi}_{_\Omega} (\vec{x'},t)] = N^{2} \int^{+\infty}_{0} d\vec{k} \int^{0}_{-\infty} d\vec{k'}  e^{(w_{k}-w_{k'})t}\Big[-J^{+} [\hat{a}(\vec{k}),\hat{b}(\vec{k'})]w_{k'} e^{\vec{k'}\cdot \vec{x'}-\vec{k}\cdot \vec{x}} \nonumber \\+ J^{-}[\hat{b}^\dag(\vec{k'}),\hat{a}^\dag(\vec{k})]w_{k}e^{\vec{k'}\cdot \vec{x}-\vec{k}\cdot \vec{x'}}\Big],
\label{fc1}
\end{eqnarray}
where the possibility of a dependence on the dissipative factors $e^{\pm\frac {\gamma t}{2}}$ has been eliminated once more. 
By using the expression (\ref{fc2}), the fundamental commutator reduces to
\begin{eqnarray}
[\hat{\Omega} (\vec{x},t), \hat{\Pi}_{_\Omega} (\vec{x'},t)]
=-N^{2}(\rho_{1}+\rho_{2})\int^{0}_{-\infty} [J^{+}e^{\vec{k'}\cdot (\vec{x'}-\vec{x})}-J^{-}e^{-\vec{k'}\cdot (\vec{x'}-\vec{x})} ]w_{k'}d\vec{k'}.
\label{pq}
\end{eqnarray}
On account of the presence of the frequency $w_{_{k}}=  \sqrt{\vec{k}\cdot \vec{k} + {\frac {\gamma ^2 }{4}}}$, this integral, and in fact all field commutators, depend sensitively on the spatial dimension, and on the interaction represented by the parameter $\gamma$. In conventional treatments, (the only) field commutator $[\hat{\Omega} , \hat{\Pi}_{_\Omega} ]$ is simply a Dirac delta function, without any trace of the background dimension, and of the interaction.
 An exact expression can be determined in the case of $(1+1)$ dimensions for the field commutator (\ref{pq}), as we shall see in the next section.

We finish this section with some remarks; first, the only nontrivial field commutator in the usual scheme is $[\psi, \Pi_{\psi}]$, and the  vanishing commutators are $[\psi, \psi^{\dagger}]$, and $[\Pi_{\psi},\Pi_{\psi}^{\dagger}]$. In the present scheme, these three commutators are necessarily nonvanishing; furthermore, the commutator $[\psi, \Pi_{\psi}^{\dagger}]$ vanishes in both schemes; therefore, there is not an isomorphism between the field commutator algebras of both schemes, and we have at hand a non-canonical quantization. The fact that we have nontrivial field commutators beyond the fundamental one, $[\psi, \Pi_{\psi}]$, suggests the presence of an underlying non-commutative field theory, intimately connected with the algebraic structure of the hyperbolic ring.
Second, although the pure commutators $[\hat{a},\hat{a}^{\dagger}]$ , and $[\hat{b},\hat{b}^{\dagger}]$ do not appear in the computation of the field commutators, they will appear below in the construction of the entropy operators (see section \ref{entropy}).

In relation to the decomposition (\ref{operator1}) we comment that in the standard scheme based on a conventional complex plane, the continuous integrals over momentum can be replaced, in the case of a finite system, by discrete sums, $\int dp \rightarrow \Sigma _{l}$; the discrete character of the momentum comes from the periodic boundary conditions imposed on the Fourier modes. However, in the formulation at hand, the solutions are not constructed in terms of plane waves as a Fourier expansion, and hence there are not periodic boundary conditions, and Fourier discrete sums  are not available on a purely hyperbolic plane description.\\

 \subsection{$(1+1)$ quantum field theory}
\begin{figure}[H]
  \begin{center}
   \includegraphics[width=.55\textwidth]{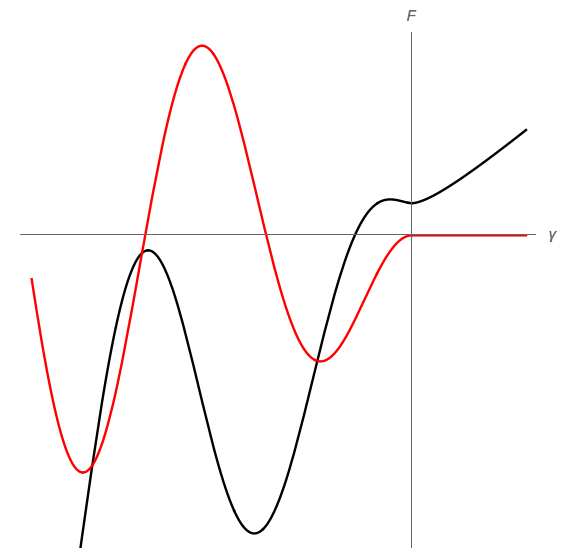}
  \caption{$F$ as a function of $\gamma$, and for a fixed $x'-x$; the black curve represents the real part; the red curve corresponds to the conventional imaginary part. Both curves are continuous in the full real interval.}  
  \label{efe1}
\end{center}
\end{figure}

\begin{figure}[H]
  \begin{center}
   \includegraphics[width=.55\textwidth]{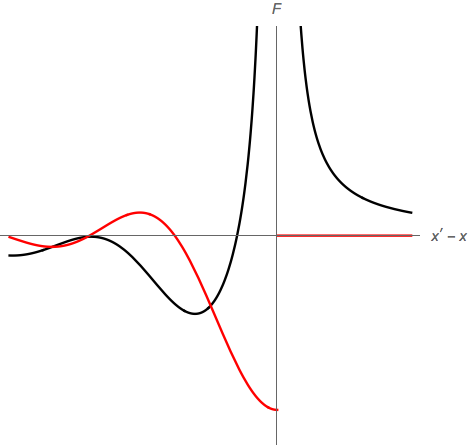}
  \caption{$F$ as a function of $x'-x$, and for a fixed $\gamma$; the black curve represents the real part; the conventional $i$-imaginary part is represented by the red curve. Both curves are no continuous at $x'-x=0$.}  
  \label{efe2}
\end{center}
\end{figure}

For the case of one spatial dimension, the following integral must be considered,
\begin{eqnarray}
\int^{0}_{-\infty} e^{\alpha k} {\sqrt{k^{2}+\beta^{2}}}dk = \frac{\pi}{2} F(\alpha,\beta),\quad F(\alpha,\beta)=\frac{\beta}{\alpha} \Big[{\rm StruveH}[1,\alpha \beta] - {\rm BesselY} [1, \alpha \beta]\Big], \quad \alpha>0 ; 
\label{struve}
\end{eqnarray}
where $BesselY$ stands for the Bessel function of the second kind ($Y_{1} (x)$) and $Struve H$ stands for the Struve function ($H_{1} (x)$); $\alpha$ must be positive for convergence.
Hence, the commutator (\ref{pq}) reduces to 
\begin{eqnarray}
[\hat{\Omega} (x,t),\hat{\Pi}_{\Omega} (x',t)] = -N^{2} (\rho_{1}+\rho_{2}) \frac{\pi}{2} \Big\{ J^{+}\frac{sgn(x'- x)+1} {2} F(x'-x,\frac{\gamma}{2})\nonumber\\+ J^{-}\frac{sgn(x'- x)-1} {2} F(x-x',\frac{\gamma}{2})\Big\};
\label {struve2}
\end{eqnarray}
the function $F(x'-x;\frac{\gamma}{2})$ is 
shown in the figure \ref{efe1} as a function of $\gamma$ and for a fixed $x'-x>0$; similarly this function is illustrated in the figure \ref{efe2} as a function of $x'-x$ for a fixed $\gamma$.
The figure \ref{efe1} shows that for $\gamma(x'-x)>0$ the function $F$
is strictly real, since the $i$-imaginary part
vanishes in this region, according to the red curve. The region with a nonvanishing 
$i$-imaginary part leads to a commutator belonging to the extended ring with two imaginary units $(i,j)$; this quantum field theory is of interest, but it will be considered elsewhere. Hence, we restrict ourselves to the case with real $F$, and thus a commutator belonging to the purely hyperbolic ring.
Furthermore, the black curve shows a nonvanishing commutator for a vanishing dissipative coe\-ffi\-cient,
\begin{eqnarray}
lim_{\gamma\rightarrow 0}F(x'-x,\frac{\gamma}{2})=\frac{2}{\pi(x'-x)},
\label{zerodiss}
\end{eqnarray}
which corresponds to a local minimum, 
and satisfies the physically admissible limits
(\ref{sgn2}); the black curve has no a global minimum.
The figure \ref{efe2} shows the function $F$ as a function of $x'-x$, for a fixed $\gamma>0$; again the $i$-imaginary part vanishes estrictly for $x'-x>0$.

Additionally
the black curve in the figure \ref{efe2} shows
the following physically admissible limits,
\begin{eqnarray}
\lim_{{x'}- {x} \rightarrow 0} [\hat{\Omega} ({x},t),\hat{\Pi} ({x'},t)] = \infty, \quad
\lim_{{x'}- {x} \rightarrow \infty} [\hat{\Omega} ({x},t),\hat{\Pi} ({x'},t)] = 0 ;
\label{sgn22}
\end{eqnarray}
which are entirely similar to those described in Eq. (\ref{sgn2}).

This new description of Dissipative CFT in $1+1$ dimensions can be identified with the quantum description of the boundary field theory in the holographic duality  $AdS_{3}/DCFT_{2}$, which has not been fully understood. In \cite{botta} this problem was investigated by considering 
an approximation for the entangled state (constructed from a canonical QFT),
in terms of the conformal ground state, dual to the BTZ spacetime. However, such a dual description breaks down for very large time, and the neccesity of a modification for the action/states is suggested. Interesting possibilities are proposed, but the non-standard description at hand of the double TFD is an obvious possibility that will be considered in  a forthcoming work; as we shall below, an entangled  state  with qualitatively different properties is in fact constructed.

\section{ Evolution operator and the grand partition function}
\label{modular}
\subsection{The Hamiltonian, and evolution operators}
\label{modular1}
The classical Hamiltonian constructed in \cite{botta} can be rewritten as
\begin{equation}
     H = \frac{1}{2}\int d^{l}x [\Pi_{\Omega}\Pi_{\bar{\Omega}} + \frac{1}{2} \partial_{i}\Omega\cdot\partial_{i}\bar{\Omega}+ \frac{\gamma}{2}j (\bar{\Omega}\Pi_{\bar{\Omega}}-\Omega\Pi_{\Omega}) - \frac{\gamma^{2}}{4}\Omega\bar{\Omega}], \label{hamilton}
\end{equation}
where each term is a $SO(1,1)$-invariant; similarly to the field commutators considered previously, each term in the Hamiltonian does not depend on the dissipative factors $e^{\pm\gamma t}$, since they are proportional to vanishing terms of the form $J^+ J^-=0$; note that $H$ is a real functional, $\bar{H}=H$. 

Thus the Hamiltonian operator has the form 
\begin{eqnarray}
\hat{H}(t;\gamma)= \frac{N^2}{2}\int^{+\infty}_{0} dk \int^{+\infty}_{0} dk'H_{kk'}
E(k,k'; \gamma, t)
\cdot \Big[ J^{+} \{ \hat{a}(k), \hat{b}(-k') \} + J^{-} \{ \hat{a}^{\dag}(k), \hat{b}^{\dag}(-k') \} \Big];
\label{modular}
\end{eqnarray}
where we have defined the following real functions,
\begin{eqnarray}
 E(k,k';\gamma,t)\equiv e^{(w_{k}-w_{k'})t} \cdot G(k,k';system),\label{E}\\
 H_{kk'}\equiv w_{k} w_{k'}-\frac{1}{2}kk'+\frac{\gamma^2}{4}-\frac{\gamma}{2}(w_{k} +w_{k'})\label{hkk}; \\
G(k,k'; system)=-\int_{system} d^{l}x \cdot e^{-\vec{x}\cdot (\vec{k}+\vec{k'})}; \label{gkk}
 \end{eqnarray} 
and we have considered that $w_{k}=w_{-k}$;
 with the basic structure $J^{+} \{ \hat{a}, \hat{b} \}+ J^{-} \{ \hat{a}, \hat{b} \}^{\dag}=J^{+} \{ \hat{a}, \hat{b} \}+h.c.$, the Hamiltonian operator is effectively Hermitian.
The function $G$ is defined
as an integral over the total system, including the subsystem $A$ and
the environment, which lie in different geometrical regions. The system can to have arbitrary shapes, and we maintain that function as arbitrary as possible; for example, the subsystem and its complement can correspond  to the two asymptotically AdS boundaries, that is typical in holographic scenarios. Furthermore, in $1+1$ dimensional (critical) systems, the subsystem A is a finite interval in an infinite total system; it can consist also of an arbitrary number of disjoint intervals \cite{cala}. In the traditional approach that invoke the usual complex unit $i$, the corresponging $G$ function is essentially a Dirac delta $\int dx e^{i x\cdot k}$, reducing the Hamiltonian operator to a single $k$-integration;
in the present scheme the function $G$ will encode the spatial configuration of the total system, which has its own physical meaning.

The purely hyperbolic QFT in construction can evolve unitarily, and the dissipation will be generated by the dynamics of the entire system;
an unitary (in the hyperbolic sense) evolution operator can be constructed as the exponential of the (Hermitian) Hamiltonian operator by using the hyperbolic unit, which in fact, it is the only one available,
\begin{eqnarray}
     e^{j\hat{H}t} \! & \equiv & \! e^{j[J^{+}\{ \hat{a}(k), \hat{b}(-k)\} + J^{-}\{ \hat{a}(k), \hat{b}(-k)\}^{\dag}]t } \nonumber \\
     \! & = & \! e^{J^{+} \{\hat{ a}(k), \hat{b}(-k) \}t }  e^{-J^{-} \{ \hat{a}(k), \hat{b}(-k) \}^{\dag}t}; \label{hermit1}\\
      \! & = & \!  J^{+}e^{ \{ \hat{ a}(k), \hat{b}(-k) \}t } + J^{-}e^{- \{ \hat{a}(k),\hat{b}(-k) \}^{\dag}t}; \label{hermit2}
 \end{eqnarray}
where the property (\ref{absorb}) has been used and the integration and real function in the expression (\ref{modular}) have been omitted by simplicity. The disentangling in the Eq. (\ref{hermit1}) is a direct consequence of the annihilation property of the basis (\ref{ortho}), in spite of the operators $\{ a,b \}$, and $\{ a,b \}^{\dag}$ do not commute; furthermore, the expression  (\ref{hermit2}) follows from the expansion of the exponentials, the annihilation property of the basis, the idempotency of the basis Eq. (\ref{idem}), and the fact that $J^{+}+J^{-}=1$.

\subsection{The charge operator,  the chemical potential, and the grand partition function}
\label{chemical}
The invariance of the Lagrangian under hyperbolic rotations implies the following conservation law,
\begin{eqnarray}
     \partial_{t} \underbrace{(\bar{\Omega}\dot{\Omega} - \Omega\dot{\bar{\Omega}} + j\gamma\Omega\bar{\Omega})}_{j_{0}} + \partial_{i} \underbrace{(\bar{\Omega}\partial^{i}\Omega
     - \Omega\partial^{i}\bar{\Omega})}_{j^{i}}\! & = & \! 0; 
     \label{car1}
\end{eqnarray}
where $j_{0}$ and $j^{i}$ are phase independent; note that $j_{0}$ has an additional $\gamma$-term respect to the usual expression for the free theory. Hence, the charge $ Q$ associated with this current will be given by,
\begin{equation}
     Q = \int d^{l} x j_{0} = \int d^{l} x (\bar{\Omega}\Pi_{\bar{\Omega}} - \Omega\Pi_{\Omega}); \label{car2}
\end{equation}
where the momenta have been defined in Eq. (\ref{momenta}). Thus, the charge operator reads
\begin{eqnarray}
     \hat{Q}(t; \gamma) \! & = & \! -\frac{j}{2} \int dx^{l} [j \{\hat{\Omega}, \hat{\Pi}_{\Omega} \} + c.c.] \nonumber \\
     \! & = & \! -N^{2}\frac{j}{2} \int^{+\infty}_{0} dk \int^{+\infty}_{0} dk' Q_{kk'} \cdot E (k,k';t) [J^{+} \{\hat{a}(k), \hat{b}(-k')\} \nonumber \\
     \! & & \! + J^{-} \{\hat{a}^{\dagger}(k), \hat{b}^{\dagger}(-k')\}], \label{car3}
\end{eqnarray}
where the function $E$ has been defined in (Eq.\ref{E}), and additionally we have defined the real function,
\begin{eqnarray}
Q_{kk'}\equiv w_{k}+w_{k'}.\label{qkk}
\end{eqnarray}
Note that the charge operator is anti-Hermitian, with the basic structure $j[J^{+} \{ \hat{a}, \hat{b} \}+ J^{-} \{ \hat{a}, \hat{b} \}^{\dag}]=J^{+} \{ \hat{a}, \hat{b} \}-h.c.$, and thus $ \hat{Q}^{\dag}=- \hat{Q}$; since both the Hamiltonian and the charge have basically the form $J^{+} \{ \hat{a}, \hat{b} \}+ J^{-} \{ \hat{a}, \hat{b} \}^{\dag}$ as operators, they trivially commute $[\hat{H},\hat{Q}]=0$, for all $t$.

In the thermal field description of a conventional charged scalar field, it is well known that exists a chemical potential $\mu$ conjugate to the quantized charge, leading to a grand partition function given by $Z(\beta,\mu)= {\rm Tr}e^{-\beta(\hat{H}-\mu\hat{Q})}$. Similarly for the case at hand, since we have constructed the corresponding conserved charge, formally the grand partition function will read,
\begin{eqnarray}
Z(\beta,\mu;\gamma)= {\rm Tr}e^{-\beta(\hat{H}-\mu\hat{Q})};
\label{gpf}
\end{eqnarray}
where $\beta=\frac{1}{T}$ is the inverse of the temperature as usual; the dependence on the dissipative parameter $\gamma$ is through the operators $\hat{H}$, and $\hat{Q}$.

\subsection{Grand partition function as a path integral}
\label{path}
In contrast to the past sections, instead of working with quantum operators we can calculate the partition functions via a path integral, which invokes necessarily the complex unit $i$; such a standard complex unit will appear only in this subsection. For systems with continuous symmetries the \emph{grand-canonical} partition function is usually the most used. Also, the imaginary-time formalism is quite useful in this approach, where you make the substitution $t \rightarrow \tau \equiv it$. If we go back to work with the real and ``imaginary'' (in the $j$-sense) parts of the field,
\begin{equation}
	\Omega = \frac{1}{2} \left( \Phi + j \Psi \right),
\end{equation}
(notice that we have introduced a factor of $1/\sqrt{2}$ with respect to (\ref{omega}) just for convenience) the expression for the partition function in four dimensions is given by \cite{lebellac, kaputsa}:
\begin{equation}
	Z = \int \mathcal{D} \Pi_\Phi \mathcal{D} \Pi_\Psi \int \mathcal{D} \Phi \mathcal{D} \Psi \exp \left( \int_0^\beta d\tau \int d^3x \left[ i \Pi_\Phi \frac{\partial \Phi}{\partial \tau} + i \Pi_\Psi \frac{\partial \Psi}{\partial \tau} - \mathcal{H} + \mu j^0 \right] \right),
\end{equation}
where $j^\mu$ is the conserved Noether current constructed in Eqs. (\ref{car1}), and (\ref{car2}), and $\mu$ the chemical potential; the $\Pi$'s represent the canonical conjugate momenta to $\Phi$ and $\Psi$ (see Eqs. (\ref{momenta})), and $\mathcal{H}$ is the Hamiltonian density (see Eq. (\ref{hamilton})). If we integrate out the momenta we are left with
\begin{equation}
	Z = \int \mathcal{D} \Phi \mathcal{D} \Psi \exp \left( \frac{1}{2} \int_0^\beta d\tau \int d^3x \left[ - \left( \frac{\partial \Phi}{\partial \tau} - i \nu \Psi \right)^2 + \left( \frac{\partial \Psi}{\partial \tau} - i \nu \Phi \right)^2 - \left( \nabla \Phi \right)^2 + \left( \nabla \Psi \right)^2 + \frac{\gamma^2}{4} \Phi^2 - \frac{\gamma^2}{4} \Psi^2 \right] \right),
\end{equation}
where $\nu \equiv - \mu + \gamma/2$, an adjusted expression for the chemical potential due to the dissipation. Notice that, apart from some minus signs coming from the hyperbolic norm, this is completely analogous the the expression for the grand partition function of a free, massive, usual complex field, where an artificial ``mass'' term, whose value is $-\gamma^2/4$, has appeared due to the dissipative interaction. It is well known \cite{lebellac, kaputsa} that the aforementioned system leads to Bose-Einstein condensation for certain value of the chemical potential, one could ask if the same is true in this case; work in this direction is currently in progress. It is nevertheless quite surprising that our system, consisting of a massless, dissipative, hyperbolic field with explicit Lorentz breaking terms, leads to such a similar partition function as a free, massive, Lorentz invariant $U(1)$ field.

The formal expressions $\mathcal{D}\Phi$ and $\mathcal{D}\Pi_{\Phi}$ appearing in the formula of the partition function, are given by $\mathcal{D}\Phi:=\Pi_{x\in\mathbb{R}^{n}}d\Phi(x)$, which  denotes something analogous to a uniform Lebesgue measure on the phase space. However, it is known that in an infinite dimensional space, a translational invariant measures cannot be properly defined. However, in the case of the scalar field, the existence of a related unique normalized measure relies on the Bochner-Minlo's theorem \cite{Gelfand}. Moreover, for an abstract commutative ring, it has been shown that a suitable outer measure can also be defined by employing a family of prime ideals of the ring \cite{Dduzik}.

\section{The vacuum as a coherent state}
\label{coherent}
A vacuum state without IR and UV divergences, and without ordering ambiguities can be defined through a coherent state (for more detail see Ref. (\cite{alex}),
\begin{equation}
      J^{+} \{ \hat{a}(k), \hat{b}(k') \}|0> = J^{+}\lambda|0>; \quad {\rm for \quad all}\quad k,k'; \quad \lambda \quad  {\rm a}\quad {\rm constant};
\label{vacuum}
\end{equation}
thus, the vacuum is  an eigenstate for a symmetrized quadratic combination of annihilation operators, the anti-commutator;      
this definition of vacuum will allow us to determine the action 
of the $J^{+}$-projection of the observables on the vacuum state, such as the Hamiltonian, and the charge operator, as we shall see below.
On the other hand, the action of the $J^{-}$-projection of observables on the vacuum state generates the following state, symmetrized in creation operators,
\begin{eqnarray}
 J^{-} \{ \hat{a}(k), \hat{b}(k') \}^{\dag}|0>= J^{-}\hat{a}^{\dag}(k) \hat{b}^{\dag}(k') |0>+J^{-}\hat{b}^{\dag}(k')\hat{a}^{\dag}(k) |0>  =J^{-}(|{b}_{k'},{a}_{k}>+ |{a}_{k},{b}_{k'}>);
 \label{excite1}
\end{eqnarray}
the notation for excited states is that the first state contains a ${\bf b}$-boson with moment $k'$, and an ${\bf a}$-boson with moment $k$, and similarly for the second state;
 since the operators $\hat{a}^{\dag}$, and $\hat{b}^{\dag}$ do not commute, the ${\bf a}$-bosons, and ${\bf b}$-bosons are distinguishable, and the two states superposed in the above expression are distinguishable. Note, however, that in the symmetrized state (\ref{excite1}), those bosons can be interchanged.

Now, the expression (\ref{vacuum}) implies that the vacuum expectation values
 for observables, in particular for the Hamiltonian and the charge, can be determined explicitly,
\begin{eqnarray}
<0|\hat{H}|0>=\frac{N^{2}}{2}<0|0>(J^+\lambda+c.c.)\int_{k}\int_{k'}H_{kk'}E, \label{vevzero1}\\
<0|\hat{Q}|0>=-\frac{N^{2}}{2}<0|0>(J^+\lambda-c.c.)\int_{k}\int_{k'}Q_{kk'}E; 
\label{vevzero2}
\end{eqnarray}
the possible divergences of these vacuum expectation values can be controlled by simply fixing
$\lambda=0$; furthermore, considering that $\lambda \equiv\lambda_{1}+j \lambda_{2}$, we have
\begin{eqnarray}
J^+\lambda+c.c. = \lambda_{1}+\lambda_{2}, \quad J^+\lambda-c.c. = j(\lambda_{1}+\lambda_{2});
\label{lambda}
\end{eqnarray}
hence, although the condition $\lambda_{1}=-\lambda_{2}$ leads to vanishing vev's, it corresponds to the case with $\lambda\sim J^-$, and thus to a vanishing trivial solution for the eigenvalue $J^+(J^-\lambda)=0$ in Eq. (\ref{vacuum}).

\section{Entangled states and pathologies cured}
\label{states}
Therefore, by considering  the definition of vacuum (\ref{vacuum}), the action of the evolution operator in its form (\ref{hermit1}) leads to an entangled two-mode state,
\begin{eqnarray}
|0(t)>\equiv
e^{j\hat{H}t}|0>=e^{\frac{N^2}{2}J^{+}\lambda\int_{k}\int_{k'}H_{kk'}E t}\cdot\exp\Big[-\frac{N^2}{2} J^{-} \int^{+\infty}_{0} dk \int^{+\infty}_{0} dk'H_{kk'}
Et \{\hat{a}^{\dagger}(k), \hat{b}^{\dagger}(-k')\} \Big]|0>;
\label{entan1}
\end{eqnarray}
Such an evolved state represents a normalized state for all $t$ provided that the original vacuum is normalized, 
\begin{eqnarray}
 <0(t)|0(t)> \equiv <0| e^{-j \hat{H}t} \cdot e^{j \hat{H}t} |0> = <0|0>.
\label{norma}
\end{eqnarray}
Furthermore, the state evolves aligned always with the original vacuum state,
\begin{eqnarray}
<0|0(t)>&=& e^{\frac{N^2}{2}(J^{+}\lambda-c.c.)\int_{k}\int_{k'}H_{kk'}E t} <0|0> \nonumber\\
&=&<0|0>\Big(cosh\Big[\frac{N^2}{2}(\lambda_{1}+\lambda_{2})\int_{k}\int_{k'}H_{kk'}E t\Big]+jsinh[...]\Big)\neq 0;
\label{aligned} 
\end {eqnarray}
where the $sinh$ function has the same argument that the $cosh$ function; in the first line of the above equation, the decomposition (\ref{hermit1}), and the (Hermitian conjugate of) definition (\ref{vacuum}) have been applied; in the second line, the Eq. (\ref{lambda}), and the expression for a purely hyperbolic phase have been considered; therefore, due to the presence 
of the $cosh$ function, the above expression does not vanish for any value of $t$ in the full interval $(-\infty,+\infty)$, in particular at the limit ${t\rightarrow +  \infty}$ (see Eqs. (\ref{foton2}), and (\ref{entangled1})); hence the time evolution does not lead out of the original Hilbert space, as opposed to the standard result of the dissipative dynamics; this result is valid in the formulation at hand independently of a finite, or infinite volume of the system of interest.

In the standard scheme, it is well known that  
the time evolution leads to out of the original space of states \cite{cele}, 
\begin{eqnarray}
lim_{t\rightarrow \infty} <0|0(t)>= lim_{t\rightarrow \infty} \exp{(-\ln cosh(\Gamma t))} =0, \label{standard1}\qquad {\rm standard}\quad {\rm scheme;}
\end {eqnarray}
strictly at finite volume.
 Such a pathology is cured in the standard QFT framework where many unitarily inequivalent representations of the canonical commutations relations are allowed, and the dissipation is interpreted at fundamental level as a tunneling between unitarily inequivalent representations \cite{cele,blasone}.
Similarly for  two different times $t, t' $, the decomposition (\ref{hermit1}) reads 
\begin{eqnarray}
e^{-j \hat{H}t'} \cdot e^{j \hat{H}t} = e^{-J^{-}\{\hat{a}^{\dagger},\hat{b}^{\dagger}\}(t-t')} \cdot  e^{J^{+}\{\hat{a},\hat{b}\}(t-t')};
\label{ns}
\end {eqnarray}
and hence
\begin{eqnarray}
<0(t')| 0 (t)> = e^{\frac{N^2}{2}(J^{+}\lambda-c.c.)(t-t')\int_{k}\int_{k'}H_{kk'}E t} <0|0> \neq 0 ; \label{ns1} 
\end {eqnarray}
where the final inequality is established using the same argument used in Eq. (\ref{aligned}), by considering the presence of a $cosh$ function.
Therefore, there are not pathologies  in the quantization realized on the pure hyperbolic ring, since we do not need invoke many unitarily inequivalent representations; the whole of the dissipative dynamics is realized by choosing only one unitarity representation of the field commutation relations.

Up to this point, we have considered that $\lambda\neq 0$; however, since the entanglement dynamics is constructed upon the $J^{-}$-direction, along which the excited states lie, we fix from this point $\lambda =0$. With this restriction and with the expansion (\ref{serie}) at hand, the state
(\ref{entan1}) can be described as 
\begin{eqnarray}
|0(t)>=e^{-J^{-}\int_{k}\int_{k'}f\{\hat{a}^{\dagger},\hat{b}^{\dagger}\}}|0>=|0>+J^{-} \sum_{n=1}^{\infty}\frac{(-1)^n}{n!} \Big(\int_{k}\int_{k'}f\{\hat{a}^{\dagger},\hat{b}^{\dagger}\}\Big)^{n}|0>\nonumber\\
=|0>-J^{-} \int_{k}\int_{k'}f(|{b}_{k'},{a}_{k}>+ |{a}_{k},{b}_{k'}>)+....;
\label{ns2}
\end{eqnarray}
where $f\equiv -\frac{N^2}{2}H_{kk'}E\cdot t$ (see Eqs. (\ref{E}), and (\ref{hkk})), and the notation (\ref{excite1}) for excited states is used; this state is then entangled in the momenta, since it can not factorized
into the product of single modes; as we shall see below, the entanglement is present even in the absence of dissipation.

\subsection{An entangled state without dissipation}
\label{wd}

The state (\ref{entan1}) described an entangled state even for a vanishing $\gamma$-parameter, with the functions (\ref{E}) and (\ref{hkk}) reduced to
\begin{equation}
     E(k,k';t, \gamma=0) = e^{(k-k')t} G(k,k';system), \quad H_{kk'}= \frac{1}{2}kk';
     \label{foton1}
\end{equation}
a distribution for a discrete time sequence such as that in Eq. (\ref{delta1}) can be identified from the expression $Et$, by identifying the expression $te^{(k-k')t}$ as the generating function for the delta function $\delta_{t}(k-k')$; hence, considering that the $k'$-integration in the state (\ref{entan1}) is convergent for $k'>0$, and for $t>0$, then we can define the sequence (\ref{delta1}) for $t=1,2,3,\ldots,$ and then in the $\lim t\rightarrow +\infty$, such a sequence satisfies (\ref{delta2}). If the system evolves in the inverse time direction $-t$, this construction is valid with an interchange of roles $k'\leftrightarrow k$. Hence, we have the following asymptotic state without dissipation
\begin{eqnarray}
     \lim_{t\rightarrow +\infty} e^{j\hat{H}t}|0> \! & = & \! \exp [-\frac{N^{2}}{4} J^{-} \lim_{t\rightarrow +\infty} \int^{+\infty}_{0} dk \int^{+\infty}_{0} dk' [te^{(k-k')t}] G(k,k')kk' \nonumber \\
     \! & & \! \cdot \{\hat{a}^{\dag}(k), \hat{b}^{\dag} (-k')\}] |0> \nonumber \\
     \! & = & \! \exp [-\frac{N^{2}}{4} J^{-} \int^{+\infty}_{0} dk G(k,k) k^2 \{ \hat{a}^{\dag}(k), \hat{b}^{\dag}(-k)\} ] |0>.
     \label{foton2}
\end{eqnarray}
From the general state (\ref{entan1}), we note that $|0(t=0)>=|0>\equiv |0>_{A}\otimes|0>_{B}$, that as a pure state, is clearly different to the case with $\gamma=0$, which evolves in time to an asymptotic state (\ref{foton2}). This fact constrasts with the standard description given in \cite{botta}, where the time dependent state $\gamma=0$ is equivalent to $t=0$, and thus the system remains in the pure state vacuum $|0>$, which 
turns out to be dual to the geometry described by two disconnected AdS
spaces. If the initial state within the scheme at hand is prepared  to be the pure state $|0(t=0)>$ dual to that geometry, then our results describe the evolution of those two copies of AdS without the dissipative interaction, going to the asymptotic state (\ref{foton2}).  In general in free quantum field theories all degrees of freedom in any region of spacetime are entangled with the degrees of freedom in other region \cite{casini}, and although the corresponding holographic realization is certainly problematic \cite{mt}, it will be addressed in forthcoming works within the present formulation.  Now we shall consider the case with dissipation.

\subsection{An asymptotic entangled state with dissipation}
\label{withd}

In similarity to the previous case, an asymptotic state for $t\rightarrow +\infty$ and with dissipation, can be determined for the general entangled state (\ref{entan1}), by considering now the function $te^{(w_{k}-w_{k'})t}$, coming from the product $Et$, as the generating function for the delta function $\delta_{t}(w_{k}-w_{k'})$, described in Eq. (\ref{delta1}) under the following considerations; first, we choice for the pair $(w_{k}, w_{k'})$, the positive roots  given by the formula (\ref{spectro2}); hence, with the construction of the discrete time sequence, we have the property (\ref{delta2}) in the form $\lim_{_{t\rightarrow +\infty}} \int^{+\infty}_{0} \delta_{t} (w_{k} -w_{k'}) f(w_{k'})dw_{k'} = f(w_{k})$. The identification of such a distribution will allow us to determine the following asymptotic state, without any restriction on the dissipative parameter,
\begin{equation}
     \lim_{t\rightarrow +\infty} e^{j\hat{H}t} |0> = \exp [-\frac{N^{2}}{4} J^{-} \int^{+\infty}_{0} dk G(k,k) \frac{w_{k}}{k}(k^{2}+\gamma^{2}-2\gamma w_{k})\{ \hat{a}^{\dag}(k), \hat{b}^{\dag}(-k)\}] \cdot |0>;
     \label{entangled1}
\end{equation}
we have considered that in the $\lim w_{k'}\rightarrow w_{k}$, also $k'\rightarrow k$. Note that this state consistently  reduces to the state (\ref{foton2}) in the $\lim \gamma\rightarrow 0$.

The dissipative interaction between the two CFT's generates an entangled state with the asymptotic state (\ref{entangled1}), which may be maximally entangled due to the dissipation; the initial state $|0(t=0)>=|0>$ can be prepared as dual to two disconnected AdS spaces, the same initial state for the free evolution described above; therefore, the free and interacting evolutions are comparable at any time, in particular   the asymptotic states (\ref{foton2}), and (\ref{entangled1}). In the standard description considered in \cite{botta}, the behavior  of the entropy suggests thermalization for later times ($ {t\rightarrow +\infty}$), and for giving a geometrical interpretation, in that reference a dissipative cosmology of scalar fields is proposed, since such a model reproduces similar equations of motion at the boundary; in order to explore the same geometrical realization for the field theory reformulated in the present work, the study of the possible thermalization of the asymptotic state  (\ref{entangled1}),  requires the construction of the entropy operators, which are considered in the next section.

Furthermore, using a standard quantization scheme, in \cite{botta} two different stages are distinguised, the early times holographically interpreted as a deformation of General Relativity, and the later times related to a wormhole, with a teleportation mechanism interchanging quantum information between the two CFT's at the boundaries. These CFT`s at the boundaries (under dissipative interaction) are those ones that we have reformulated on the hyperbolic plane in the scheme at hand, and thus the dissipation at one boundary is understood as an exchange of energy with the other boundary.
However, the holographic dictionary has not be fully established in a satisfactory way for this system, and it is suggested that a slight modification of action/states of the double CFT system will allow to capture the physics on the gravity side; this issue deserves further investigation within the noncanonical quantization framework at hand, and it will be considered in future explorations (see  section \ref{cr}).

\section{Entanglement entropy}
\label{entropy}
Within the standard treatments the introduction of  the concept of entanglement entropy in dissipative dynamics is based on the {\it factorization} of the operator that determines the evolved vacuum,
\begin{eqnarray}
|0(t)>=e^{-\hat{S}_{A}}|I>=e^{-\hat{S}_{B}}|I>, \quad |I>=e^{\sum_{k}\hat{A}^{\dagger}_{k}\hat{B}^{\dagger}_{k}}|0>;
\label{factor}
\end{eqnarray}
where $I$ stands for {\it interaction}, and the operators $S$ depend only on one type of creation/annihilation operators, and on the time, $\hat{S}_{A}=\hat{S}_{A}(\hat{A},\hat{A}^{\dagger};t)$, and similarly for $S_{B}$ (see \cite{botta},\cite{cele}, and \cite{blasone} for more details). In this form, the evolved vacuum is expressed in terms of only one subsystem, which is considered as an "open" one; moreover, the expectation values of these operators at the evolved vacuum, say $S_{A}$ for the subsystem of interest, is connected directly with the density operator of the total system, $\rho=|0(t)><0(t)|$, from which the reduced density operator is obtained by tracing out the $B$-degrees of freedom, $\rho_{A}=Tr_{B}\rho$, and then $\rho_{A}(t)=e^{-S_{A}(t)}$; this is basically the justification of the formal interpretation of these operators as the entropy of the subsystems.
In the scheme at hand, we follow the same strategy of rewriting the evolved vacuum in the factored form (\ref{factor}), taking into the account that we shall be strongly restricted by the algebraic structure of the ring; it is in this sense that the final expressions obtained for the subsystems, namely, Eqs. (\ref{entroA}), and (\ref{entroB}), must be interpreted as entropy operators. However, the construction of a density operator and its traces on the hyperbolic ring deserves more developments, and we are addressing this problem in \cite{rcf}; here, we restrict ourselves to the construction of certain operators that can be interpreted as entropy operators, and we show that they effectively drive the dissipative dynamics, and can be useful as measure of entanglement.

\subsection{The subsystem A}
Consider the following commutator constructed using the commutator (\ref{fc2}), and an additional commutator, $[\hat{a}_{k}, \hat{a}^{\dag}_{k'}] = \varepsilon\delta^{+}(k-k')$;
\begin{eqnarray}
\! & & \! J^{-} \Big[ \int_{k}\int_{k'} f(k,k') \{ \hat{a}^{\dag}_{k}, \hat{b}^{\dag}_{-k'}\},\quad \int_{q_{1}}\int_{q_{2}} g(q_{1},q_{2}) \hat{a}^{\dag}_{q_{1}} \hat{a}_{q_{2}}\Big] \nonumber \\
\! & = & \! 2J^{-}\bar{\rho} \int_{k}\int_{q_{2}}\Big( \int_{k'} f(k,k')\cdot g(-k',q_{2})\Big) \hat{a}^{\dag}_{k}\hat{a}_{q_{2}} - 2 J^{-}\varepsilon \int_{k'}\int_{q}\Big( \int_{k} f(k,k')\cdot g(q,k)\Big) \hat{a}^{\dag}_{q}\hat{b}^{\dag}_{-k'},
\label{fg}
\end{eqnarray}
where we have used our fundamental commutator, $[\hat{a}_{q}, \hat{b}_{q'}]=\rho\delta^{+}(q-q')$, and we have defined again $f\equiv -\frac{N^2}{2}H_{kk'}E\cdot t$, using Eqs. (\ref{E}), (\ref{hkk}), (\ref{gkk}), and considering that Eq. (\ref{entan1}) describes the evolved vacuum in terms of the function $f$. For simplicity, in the above expression we only display the relevant dependence on $(k,k')$ for the functions $f$, and $g$; 
all $k$-integration intervals are $(0,+\infty)$.
Now this commutator closes on the element $J^{-}\int_{q_{1}}\int_{q_{2}} g \hat{a}^{\dag}_{q_{1}}\hat{a}_{q_{2}}$, provided that the following conditions are satisfied
\begin{equation}
\varepsilon = 0, \qquad \int_{k'} f(k,k') \cdot g(-k',q) dk'= g(k,q);
\label{fg1}
\end{equation}
the first condition trivializes the usual commutator $[a,a^{\dag}]=0$, the second condition represents an integral equation for the function $g$.
The presence of this integral restriction is a consequence of the absence of the conventional complex unit $i$, especifically through the absence of a Fourier representation for the Dirac delta function; this absence is the final responsable of the appearance of double integrations in our expressions, in contrast with those ones within a conventional treatment.
Thus, the commutator (\ref{fg}) is reduced to the form $[X,Y]=2\bar{\rho}Y$, with the appropriate identification.

For this special commutator we can use the braiding identity $e^{X} \cdot e^{Y}= e^{e^{2\bar{\rho}}\cdot Y} \cdot e^{X}$ (see \cite{brunt} for special cases of closed forms of the Baker-Camphell-Hausdorff formula), and to obtain,
\begin{equation}
e^{J^{-}\int_{k}\int_{k'} f\{\hat{a}^{\dag}_{k}, \hat{b}^{\dag}_{-k'}\} } \cdot e^{J^{-}\int_{q_{1}}\int_{q_{2}} g\hat{a}^{\dag}_{q_{1}} \hat{a}_{q_{2}}} = e^{J^{-} e^{2\bar{\rho}}\cdot \int_{q_{1}}\int_{q_{2}} g\hat{a}^{\dag}_{q_{1}} \hat{a}_{q_{2}}} \cdot e^{J^{-}\int_{k}\int_{k'} f\{\hat{a}^{\dag}_{k}, \hat{b}^{\dag}_{-k'}\} };
\label{fg2}
\end{equation}
hence, the exponential operators commute modulo a nontrivial dilation exponential $e^{2\tilde{\rho}}$; the braiding identity will have the restriction $\rho \neq 2in\pi$, when formulated on a conventional complex plane, and hence an unnecessary restriction on the purely hyperbolic plane used here. Therefore, in the well defined $\lim \rho\rightarrow 0$, such operators strictly will commute. The above expression can be rewritten as
\begin{equation}
e^{J^{-}\int_{k}\int_{k'} f\{\hat{a}^{\dag}_{k}, \hat{b}^{\dag}_{-k'}\} } \cdot e^{J^{-}\int_{q_{1}}\int_{q_{2}} g\hat{a}^{\dag}_{q_{1}} \hat{a}_{q_{2}}} \cdot e^{-J^{-}\int_{k}\int_{k'} f\{\hat{a}^{\dag}_{k}, \hat{b}^{\dag}_{-k'}\} }   = e^{J^{-} e^{2\bar{\rho}}\cdot \int_{q_{1}}\int_{q_{2}} g\hat{a}^{\dag}_{q_{1}} \hat{a}_{q_{2}}};
\label{fg22}
\end{equation}
which is understood as an adjoint dilation for $e^{J^{-}\int_{q_{1}}\int_{q_{2}} g \hat{a}^{\dag}_{q_{1}}\hat{a}_{q_{2}}}$.
These expressions allow us to consider again the definition of the vacuum state, and we complement the previous definition (\ref{vacuum}), with an additional coherent state condition,
\begin{eqnarray}
\hat{a}^{\dag}_{q}\hat{a}_{q'} |0> = \alpha|0>,\quad \rightarrow \quad
e^{J^{-}\int_{q_{1}}\int_{q_{2}}g\hat{a}^{\dag}_{q_{1}}\hat{a}_{q_{2}}} |0> = e^{J^{-}\alpha\int_{q_{1}}\int_{q_{2}}g(q_{1},q_{2})}|0>, \label{fg3}
\end{eqnarray}
where $\alpha$ is a real constant, and the double integration on the function $g$ is determined by the integral equation (\ref{fg1}), explicitly we have,
\begin{equation}
\int_{k}\int_{q}\int_{k'} f(k,k';\gamma, t) \cdot g(-k',q;\gamma,t) = \int_{k}\int_{q}g(k,q;\gamma,t);
\label{fg33}
\end{equation}
thus, such an integration will reduce to an expression that depends on 
the spatial configuration of the (sub)system through the $G$ function, the dissipation parameter $\gamma$, and the time $t$. Furthermore,
the complete action of the operator (\ref{fg2}) reads
\begin{equation}
e^{J^{-}\alpha\int_{k}\int_{k'} g(k,k')}\cdot
e^{J^{-}\int_{k}\int_{k'} f\{\hat{a}^{\dag}_{k}, \hat{b}^{\dag}_{-k'}\} } |0> = e^{J^{-} e^{2\bar{\rho}}\cdot \int_{q_{1}}\int_{q_{2}} g\hat{a}^{\dag}_{q_{1}} \hat{a}_{q_{2}}}\cdot e^{J^{-}\int_{k}\int_{k'} f\{\hat{a}^{\dag}_{k}, \hat{b}^{\dag}_{-k'}\} } |0>;
\label{fg4}
\end{equation}
the left hand side reproduces basically the expression (\ref{entan1}) for the evolved vacuum, since the construction at hand has been made for this purpose. Furthermore, on the right side, we have exactly the same evolved
state, but disentangled from an operator that involves only $a$-degrees of freedom; this purely $a$-type operator will correspond to the entropy operator for the subsystem $A$;  thus, the above expression has the form 
\begin{eqnarray}
 S(A)|0(t)> =e^{J^{-}\alpha\int_{k}\int_{k'} g(k,k')} |0(t)>,\quad 
S(A)\equiv e^{J^{-} e^{2\bar{\rho}}\cdot \int_{q_{1}}\int_{q_{2}} g(q_{1},q_{2})\hat{a}^{\dag}_{q_{1}} \hat{a}_{q_{2}}};
\label{entroA}
\end{eqnarray}
hence
\begin{eqnarray}
<0(t)|S(A)^{\dagger}S(A)|0(t)>= e^{\alpha\int_{k}\int_{k'} g(k,k')}<0|0>
;\quad 
<0(t)|S(A)|0(t)>= e^{J^{-}\alpha\int_{k}\int_{k'} g(k,k')}<0|0>;
\label{sa}
\end{eqnarray}
where strictly $g=g(k,k'; t, \gamma)$; the expectation values at the evolved vacuum for $S(A)$, follows directly from Eqs. (\ref{entroA}), (\ref{norma}); note that the second expression in Eq.(\ref{sa}) is of the form (\ref{projector}),
and hence its norm is basically given by $e^{\frac{1}{2}\alpha\int_{k}\int_{k'} g(k,k')} $. At the end the expectation values
are determined by the eigenvalue of the coherent state condition (\ref{fg3}), and 
by the solution to the integral equation (\ref{fg1}), which will be addressed in forthcoming works. From this solution one expects the increasing of the entropy of the subsystem, expressed by the monotonical increasing of the expectation values in the full interval $t\in(0,\infty)$. This requirement is satisfied through the constraint $\alpha\int_{k}\int_{k'} g(k,k')>0$,  for the full time interval, which must be considered in the search of solutions for that integral equation.
 Furthermore, this requirement on the field theory side, must be consistent with the concave character of the entanglement entropy respect to geometric parameters on the gravity side, which ensures strong subadditivity (see concluding remarks).\\

\subsection{The subsystem B: the environment}
\label{subB}
Along the same lines described in the previous subsection, we consider now the commutator
\begin{eqnarray}
\! & & \! J^{-} \Big[ \int_{k}\int_{k'} f(k,k') \{ \hat{a}^{\dag}_{k}, \hat{b}^{\dag}_{-k'}\},\quad \int_{q_{1}}\int_{q_{2}} h(q_{1},q_{2}) \hat{b}^{\dag}_{-q_{1}} \hat{b}_{q_{2}}\Big] \nonumber \\
\! & = & \! -2J^{-}\bar{\rho} \int_{k'}\int_{q}\Big( \int_{q'} f(-q',k')\cdot h(q',q)\Big) \hat{b}^{\dag}_{-k'}\hat{b}_{q} - 2 J^{-}\varrho \int_{k}\int_{q'}\Big( \int_{k'} f(k,k')\cdot h(q',-k')\Big)\hat{b}^{\dag}_{-q'} \hat{a}^{\dag}_{k},\nonumber \\
\label{fh}
\end{eqnarray}
where $[\hat{b}_{k}, \hat{b}^{\dag}_{k'}] = \varrho\delta^{+}(k-k')$, and the function $h(k,k')$ must be determined; again, all $k$-integration intervals are $(0,+\infty)$.
The commutator will take the closed form $[X,Y]=-2\bar{\rho}Y$, under the restrictions
\begin{eqnarray}
\varrho=0, \quad \int_{q'} f(-q',k')\cdot h(q',q)=h(k',q);
\label{fh1}
\end{eqnarray}
which must be compared with the restrictions (\ref{fg1}), and we realize that in general the unknown functions $g$, and $h$ are different to each other, although certainly the integral equations are similar to each other. Furthermore, we have a coherent state condition similar to the expression (\ref{fg3}), 
\begin{eqnarray}
\hat{b}^{\dag}_{q}\hat{b}_{q'} |0> = \beta|0>,\quad \rightarrow \quad
e^{J^{-}\int_{q_{1}}\int_{q_{2}}h\hat{b}^{\dag}_{q_{1}}\hat{b}_{q_{2}}} |0> = e^{J^{-}\beta\int_{q_{1}}\int_{q_{2}}h(q_{1},q_{2})}|0>, \label{fg2B}
\end{eqnarray}
where $\beta$ is a real constant; similarly to the Eq. (\ref{fg4}), 
one obtains a disentangled expression for the entropy operator for the subsystem B, 
\begin{eqnarray}
 S(B)|0(t)> =e^{J^{-}\beta\int_{k}\int_{k'} h(k,k')} |0(t)>,\quad
S(B)\equiv e^{J^{-} e^{-2\bar{\rho}}\cdot \int_{q_{1}}\int_{q_{2}} h(q_{1},q_{2})\hat{b}^{\dag}_{q_{1}} \hat{b}_{q_{2}}};
\label{entroB}
\end{eqnarray}
hence
\begin{eqnarray}
<0(t)|S(B)^{\dagger}S(B)|0(t)>= e^{\beta\int_{k}\int_{k'} h(k,k')}
;\quad 
<0(t)|S(B)|0(t)>= e^{J^{-}\beta\int_{k}\int_{k'} h(k,k')};
\label{sb}
\end{eqnarray}
which can be obtained from $S(A)$ in Eq. (\ref{sa}) with the transformation
\begin{eqnarray}
g \rightarrow h, \quad \rho \rightarrow -\rho, \quad \alpha \rightarrow \beta, \quad \hat{a}\rightarrow \hat{b}.
\end{eqnarray}

Note that although the entropy operators $S(A)$, and $S(B)$ involve only $a$-type, and $b$-type modes respectively, both operators depend on the quantity $\rho=[\hat{a},\hat{b}]$, which measures a nontrivial correlation between those modes. 
Furthermore, the equations (\ref{entroA}), and (\ref{entroB}), imply additionally that
\begin{eqnarray}
\Big( S(A)-S(B)\Big)|0(t)>=\Big(e^{J^{-}\alpha\int_{k}\int_{k'} g(k,k')} - e^{J^{-}\beta\int_{k}\int_{k'} h(k,k')}\Big)|0(t)>,
\label{zero}
\end{eqnarray}
thus, the conserved entropy for the complete system can be obtained in particular from the condition $\alpha=\beta$, and assuming that we can find solutions such that $g=h$ for any time; however, these particular restrictions are unnecessary, since the entropy of a closed physical system (the complete system), should never decrease, and one expects rather subadditive inequalities from this equation, which will encode the irreversibility of entropy (see concluding remarks). A detailed analysis requires to know explicitly the functions $g$ and $
h$; 
technically the finding of explicit solutions for the integral equations (\ref{fg1}), (\ref{fh1}), requires to specify first the system of interest, which will allow us to determine the explicit form for the function $G$ that depends sensitively on the spatial configuration of the system; then, the function $f$ takes an specific form and the solutions can be in principle determined.

\section{Concluding remarks and perspectives}\label{conlusion}
\subsection{Thermal field theory}
\label{tft}
With the operators $S(A)$, and $S(B)$, controlling the time evolution for the separate subsystems,
we may to be in conditions of introducing formally the temperature $T(t)$ of the system through of the concept of {\it free energy} functional; in the traditional description of the quantum dissipation given in \cite{cele}, the Hamiltonian of the system has the form ${\cal{H}}= {\cal{H}}_{0}+{\cal{H}}_{I}$, where basically ${\cal{H}}_{0}\sim \hat{a}^{\dagger}\hat{a}-\hat{b}^{\dagger}\hat{b}$, and ${\cal{H}}_{I} \sim \hat{a}^{\dagger}\hat{b}^{\dagger}-\hat{a}\hat{b}$, and the functional for the $a$-modes reads,
\begin{eqnarray}
{\cal F}_{A}=<0(t)|{\cal H}_{a}-kT(t)S(A)|0(t)>,
\nonumber
\end{eqnarray}
where ${\cal H}_{a}\sim \hat{a}^{\dagger}\hat{a}$, and the interaction part ${\cal H}_{I}$ is not involved. However, as we have seen before,
the Hamiltonian in the approach at hand can not be splitted into the form
${\cal{H}}_{0}+{\cal{H}}_{I}$, and it has rather the form of the interaction Hamiltonian in the usual scheme, ${\cal H}_{total}\sim J^{+} \{ \hat{a}, \hat{b} \}+ J^{-} \{ \hat{a}^{\dagger}, \hat{b} ^{\dagger}\}$, which does not admit a decomposition into $a$-modes plus $b$-modes. Thus, in the present scheme one may propose that
\begin{eqnarray}
{\cal F}_{A}=<0(t)|{\cal H}_{total}-kT_{A}(t)S(A)|0(t)>,
\nonumber
\end{eqnarray}
and similarly for the subsystem B; from these expressions for the functionals, the Bose distribution for the activated modes, adiabatic variations of temperature, the first principle of thermodynamics of the system interacting with the environment at constant temperature, boson condensation, etc, must be in principle obtained; more investigation is clearly needed.

\subsection{Holographic perspectives} 
\label{cr}

As emphasized in \cite{mt}, it is not clear whether a system of two interacting CFT's admits a holographic description, and an attempt in made in \cite{botta} invoking a canonical quantization scheme for the dissipation, and remarking two peculiarities for such a system: a time-dependent entropy and the breaking of the Lorentz symmetry by dissipative effects. The approach at hand allows to reveal two additional peculiarities intimately connected, first the emergence of an internal Lorentz symmetry, indistinctly whether the dissipation is turned on or turned off, and second, a non-canonical quantization framework committed with this symmetry. This new symmetry revealed, can play the key role in the holographic realization of both free and interacting CFT's, and it will be the subject of forthcoming works.

It has been shown that subadditivity of entanglement entropy is satisfied
for several explicit examples, once the holographic computations have been made \cite{hirata}; subadditivity requires that the entropy is a concave function respect to the geometric parameters; additionally the (covariant) entropy bound known as the Bousso bound \cite{bou} is saturated in the case of CFT's in absence of dissipation; is an open problem to explore subadditivity and Bousso bounds in dissipative CFT's within the noncanonical quantization scheme developed here.

A basic feature of the holographic correspondence is to relate the microscopic information on the quantum field theory side, to the geometric 
quantities on the gravity side; the Ryu-Takayanagi proposal stablishes that the area of a minimal surface in a holographic geometry is related to the entanglement between the degrees of freedom contained in that region and the degrees of freedom of its complement \cite{ryu}; this proposal has been tested positively in many forms, from  scenarios with particular symmetries \cite{huerta}, to higher derivative gravity theories \cite{miao}; the possible quantum corrections to this area law formula has been also considered \cite{har,mal}.
This conjecture allows us to stablish, within the scheme at hand, that
\begin{eqnarray}
<0(t)|S(A)|0(t)>=\frac{1}{4G_{N}} Area,
\nonumber
\end{eqnarray}
where the entropy on the left hand side has been determined in principle from the quantum states of the system; since there exist only few techniques for calculating the entanglement entropy for interacting field theories, the area law provides a geometrical perpective based on minimal surfaces; additionally, assuming that this law is independent on the quantization scheme, we are providing a nontrivial novel ingredient, namely, a noncanonical QFT perspective for the left hand side of that formula, in order for testing and exploring the conjecture in different holographic scenarios. In particular we have at hand an alternative approach for studying the behavior of the entanglement entropy as a time dependent function for interacting field theories, an open problem nowadays.\\

\subsection{Highlights}

Finally, we describe here, as concluding remarks, a list of highlights of this paper,
which are independent on the possible holographic applications of the results:

\hspace{1cm} 1.- A new perspective for quantizing a field theory, beyond the standard and well-established harmonic oscillator physics.

\hspace{1cm}2.-  To quantize a field theory using a non-standard complex unit, which is motivated by revealing a new symmetry  in the Schwinger-Keldysh formulation.

\hspace{1cm}3.- This quantization scheme has cured pathologies coming from the standard harmonic oscillator quantization, which can be extended to more general cases for open systems, for example those systems involving gauge fields and fermionic fields.

\hspace{1cm}4.- The new continuous symmetry implies a qualitative change in the construction of the corresponding partition function, which  includes now the term corresponding to the charge conserved; this new term is not considered in the previous treatments of the existing model.

\hspace{1cm}5.- Quantum observables, operators such as the entropy operators, etc, show qualitative and quantitative new properties;
additionally the field commutators show evidence of an underlying non-commutative field theory;  the algebra of the commutation relations is not isomorphic to the algebra defined by the damping harmonic oscillator.  

\hspace{1cm}6.-Furthermore, the field commutators show a nontrivial dependence on the dissipative parameter (interaction); this fact is explicitly developed for $1+1$ QFT; similarly the field commutators depend sensitively on the background dimension. These features are absent 
in all previous treatments on the subject, where the field commutators are basically  Dirac delta functions in all dimensions, such as a free QFT.

\hspace{1cm}7.- A new description of thermal field theory described above in the subsection \ref{tft}.

There is no exists, to the knowledge of the authors, previous treatments on the Schwinger-Keldysh formulation, that include any of  these new elements.\\

{\bf Acknowledgements:}
This work was supported by the Sistema Nacional de Investigadores (M\'exico) and by CONACyT grant CB-2017-283838. The graphics have been made using Mathematica.

\end{document}